\documentclass[12pt]{article}
\usepackage{fullpage}
\usepackage{graphicx}
\usepackage{amssymb}
\usepackage{amsmath}
\usepackage{multicol}

\newcommand{\Fig}{Figure }
\newcommand{\pr}{\partial{}}
\newcommand{\bphi}{\mbox{\boldmath $\phi$}}
\newcommand{\bx}{\mbox{\boldmath $x$}}

\newcommand{\news}{\setcounter{equation}{0}\quad}
\title{
\vskip 2cm Baby Skyrmion chains}
\author{David Foster\\[10pt]
{\em \normalsize Department of Mathematical Sciences,
Durham University, Durham DH1 3LE, U.K.}\\[10pt]
{\normalsize Email: d.j.foster@durham.ac.uk}
}

\begin{document}

\maketitle

\begin{abstract}
Previous results on multi-charged Baby Skyrmion solutions have pointed to a modular structure, comprised of charge two rings and single charge one Skyrmions, which combine to form higher charged structures. In this paper we present numerical evidence which shows an alternative finite chain, multi-charged global energy minimum Baby Skymion solution. We then proceed from the infinite plane, to Baby Skyrmions on a cylinder and then a torus, to obtain the solutions of periodic Baby Skyrmions, of which periodic segments will correspond to sections of large charge Baby Skyrmions in the plane.
\end{abstract}

\newpage

\section{Introduction}\news
The Skyrme model is a $(3+1)$ dimensional non-linear theory of pions and admits topological soliton solutions which describe baryons. In this paper we are concerned with a lower dimension analogue, known as the Baby Skyrmion model \cite{BW1}. This theory has the Lagrangian density

\begin{equation}
\mathcal{L} =  \frac{1}{2}\partial_\mu\bphi\cdot\partial^\mu\bphi -\frac{\kappa^2}{4}(\partial_\mu\bphi\times\partial_\nu\bphi)\cdot
(\partial^\mu\bphi\times\partial^\nu\bphi) -m^2(1-\phi_3), \label{L}
\end{equation}
where the field is a three component scalar, $\bphi = ( \phi_1,\phi_2,\phi_3)$ and takes its value on the unit sphere $\bphi \cdot \bphi =1$.

 Theory \eqref{L} is a modified $O(3)$ $\sigma$-model which includes a fourth power derivative and a symmetry breaking potential. The fourth power derivative is commonly known as the Skyrme term. The Skyrme term coupled with the potential term gives rise to a scale in \eqref{L}, and therefore by Derrick's theorem \cite{Derrick}, the theory admits soliton solutions. The potential is analogous to the pion mass term in the $(3+1)$ dimensional Skyrme model. The positive constant $\kappa$ pre-multiplying the Skyrme term, in combination with the mass, $m$, determines the size of the Baby Skyrmion as being proportional to $\sqrt{\kappa/m}$.
For finite energy, $\bphi$ has to be a constant at the spacial boundary, thus we require the condition
\begin{equation}
\lim_{|\bx| \to \infty} \bphi = (0,0,1).
\end{equation}
The physical space, is a one-point compactification $\mathbb{R}^2 \cup \{\infty\}$, which is topologically equivalent to $S^2$. This compactification of the physical space and the unitary nature of the target space, gives rise to the field configuration $\bphi$ at fixed time being a map
\begin{equation}
 \bphi: S^2 \to S^2.
\end{equation}
Such maps can be classified by the homotopy class $ \pi_2(S^2) = \mathbb{Z}$, which gives the theory a topological degree, otherwise known as a topological charge, $B$. The topological charge of the map $\bphi$, is found to be the pull back of an area form of the target $S^2$, integrated over the physical space

\begin{equation}
 B = \deg [\bphi] = \frac{1}{4 \pi} \int_{\mathbb{R}^2} \bphi \cdot (\pr_1 \bphi \times \pr_2 \bphi) d^2x. \label{Q}
\end{equation}

The aim of this paper is to discuss the structure of static multi-charge solutions, focusing on the chain-like structure that we conjecture to be the minimum energy configuration of multi-charge Baby Skyrmions. Firstly, we shall outline the well known symmetry reduction used to calculate charge one and two Baby Skyrmions, then in more detail discuss the numerical computation of higher charge structures, producing the minimum energy chain configuration. We then extend our investigation to include periodic spaces, firstly investigating infinite chains on $\mathbb{R}^1 \times S^1$ and then Baby Skymion crystal structures on the torus, $ \mathbb{T}^2$.

\section{Charge one and two}\news
Static Baby Skyrmions are energetic minima of the energy associated with \eqref{L}, namely,
\begin{equation}
 E = \int _{\mathbb{R}^2} \left(\frac{1}{2} \pr_i \bphi \cdot \pr_i \bphi + \frac{\kappa^2}{2} \left| \pr_1 \bphi \times \pr_2 \bphi\right|^2 + m^2(1-\phi_3)\right)d^2x. \label{E}
\end{equation}
The static field equation found from the variation of \eqref{E} is found to be a highly non-linear partial differential equation, 
\begin{eqnarray}\label{phi-eq}
0&=& \partial{}_i \partial{}_i \bphi + m^2 \mathbf{e}_3 + \lambda \bphi \\ \nonumber
&+&  \kappa^2 \left(\partial{}_i\partial_i  \bphi (\partial{}_j \bphi \cdotp \partial{}_j \bphi) + \partial{}_i \bphi (\partial{}_i\partial{}_j \bphi \cdotp \partial{}_j \bphi) - \nonumber \partial{}_i \partial{}_j \bphi (\partial{}_i \bphi \cdotp \partial{}_j \bphi) - \partial{}_j \bphi (\partial{}_i \partial{}_i \bphi \cdotp \partial{}_j \bphi)\right), \nonumber
\end{eqnarray} 
where $\mathbf{e}_3 = (0,0,1)$ and $\lambda$ is a suitable Lagrange multiplier to impose the condition $\bphi \cdot \bphi =1$.

The only way to solve for $\bphi$ is to use a numerical technique. Throughout this paper we used gradient flow to solve these partial differential equations.
A very effective simplification of \eqref{E}, and hence its corresponding field equation \eqref{phi-eq}, is to take advantage of the underlying symmetry of an $O(2)$ rotation in the physical space, which can then be neutralised by an iso-rotation of the target space. This technique is extensively covered in \cite{BW1}, where it is shown that the field, $\bphi$, can be re-expressed with axial symmetry in terms of a profile function $f(r)$, and a polar angle $\theta$ as
\begin{equation}
 \bphi(\bx)=(\sin f \cos(B\theta-\chi), \sin f \sin(B\theta-\chi),\cos f ), \label{iso-anz}
\end{equation}

with a global phase $\chi \in [0,2 \pi )$. The profile function $f(r)$ takes the values $f(0) = \pi$ and $f(\infty)=0$. Using this simplification the static energy \eqref{E} becomes,

\begin{equation}
 E =2 \pi \int_0^{\infty}  \left( \frac{1}{2} {f'}^2 +(1+ \kappa^2 {f'}^2)\frac{B^2}{2 r^2} \sin^2 f + m^2(1-\cos f) \right)r dr . \label{Ef}
\end{equation}

\begin{figure}[!ht]
 \centering
 \includegraphics[width = 11cm]{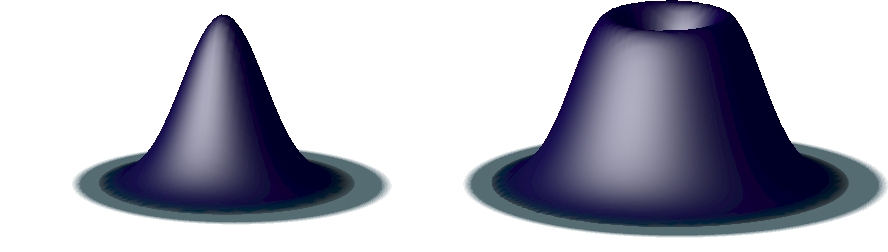}
\caption{
Two-dimensional plots of the topological charge density of Baby Skyrmions
with $B=1$ on the left and $B=2$ on the right.
}
\label{fig-2}

\end{figure}

Varying \eqref{Ef} gives
\begin{equation}
  f''\left(r+ \frac{\kappa^2 B^2 \sin^2 f}{r}\right)+ f'\left(1-\frac{\kappa^2 B^2}{r^2} \sin^2 f +f' \frac{\kappa^2 B^2}{2r} \sin(2f)\right)- \frac{B^2}{2r} \sin(2f) - m^2 r \sin f  = 0. \label{profile_file}
\end{equation}
For comparative consistency with \cite{BW1}, we chose $\kappa =1 $ and $m^2 =0.1$. A gradient flow algorithm is then implemented using \eqref{profile_file} to minimise \eqref{Ef}. The minimum energies for $B=1$ and $B=2$ are found to be $E_1 =19.66$ and $E_2 = 36.90$ respectively. These energies are within $0.03\%$ agreement with the results of \cite{BW1}. The profile functions for these solutions are then used to generate the charge density plots in figure~\ref{fig-2}.

Only $B=1,2$ minimum energy Baby Skyrmion solutions can be calculated using the above method, due to the symmetry imposed by \eqref{iso-anz}. For $B>2$ this approach creates ring Baby Skymion structures, which are not stable to perturbations which break the axial symmetry. Therefore, to calculate $B>2$ minimum energy Baby Skyrmions, full field two-dimensional gradient flow is required.

\section{Larger charge Baby Skyrmions $B>2$}\news
As stated above, for $B >2$ it is found \cite{BW1} that the axially symmetric solutions are not stable to axial breaking perturbations in full field simulations. To proceed we solve the static field equation \eqref{phi-eq}, using gradient flow. For initial conditions it is conducive to use stereographic coordinates
\begin{equation}
 W = \frac{\phi_1 + i \phi_2}{1+\phi_3}.
\end{equation}

 A sum of $n$ charge one Baby Skyrmions in stereographic coordinates, equally separated on a circle with arbitrary phase is used as the initial condition. This configuration is then numerically minimised using gradient flow on a $200 \times 200$ lattice with spacing of $\Delta x =0.2$. As a comparison to validate the two dimensional minimisation, the minimum energies of the $B=1$ and $B=2$ are calculated to be $E_1=19.65, E_2 = 36.90$. These $B=1,2$ values differ slightly from the two dimensional results in \cite{BW1},\cite{TW1} and \cite{CP1}, the latter results seem to differ by $1\%$  compared to the greater accuracy axial solutions. Our results are in better agreement with the previous dimensionally reduced solution with $E_1=19.66$, and have a closer agreement with the dimensionally reduced results of the collective literature. We believe our slightly greater accuracy is due to the use of fourth-order accurate finite differences in comparison to the second-order differences commonly used.

\subsection{$B=3,4$}

\begin{figure}[!ht]
 \centering
 \includegraphics[width = 11cm]{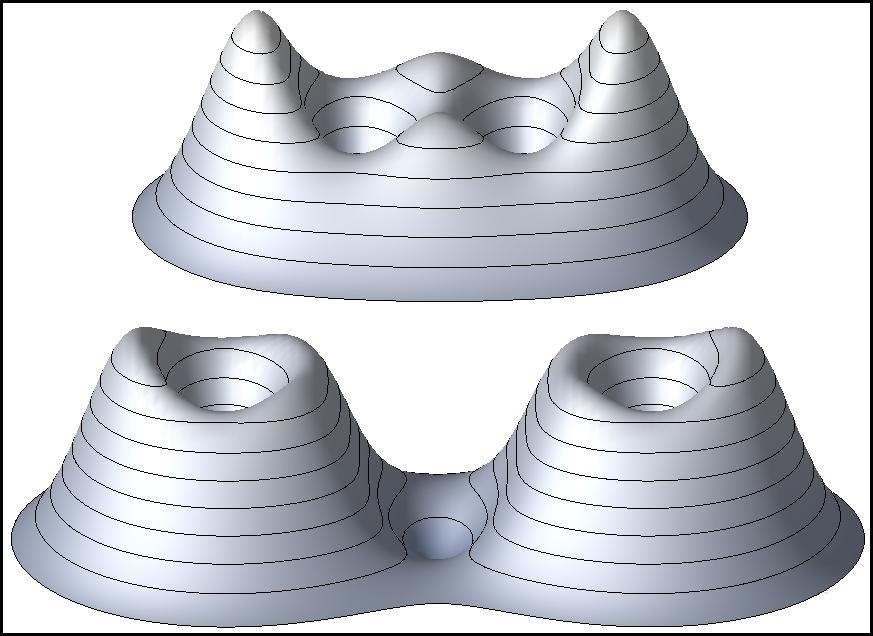}
\caption{
Two-dimensional plots of the topological charge densities for the $B=3$ Baby Skyrmion on the top, and the $B=4$ Baby Skyrmion on the bottom.
}
\label{Q3-Q4-comparison}

\end{figure}

Minimisation of $B=3,4$ initial conditions reveal energies of $E_3 = 55.58,E_4=73.61$, which are close to the previous literatures energies plus a $1\%$ perceived discrepancy in their computations. We also gain similar charge density distributions as in \cite{BW1}, shown in figure~\ref{Q3-Q4-comparison}. 

The configurations for $B=1,2,3,4$ which we obtain are essentially identical to those of \cite{BW1}. The first major difference between our results and the results contained in \cite{BW1} arises at $B=5$.
\newpage

\subsection{$B=5$}
\begin{figure}[!ht]
 \centering
 \includegraphics[width = 11cm]{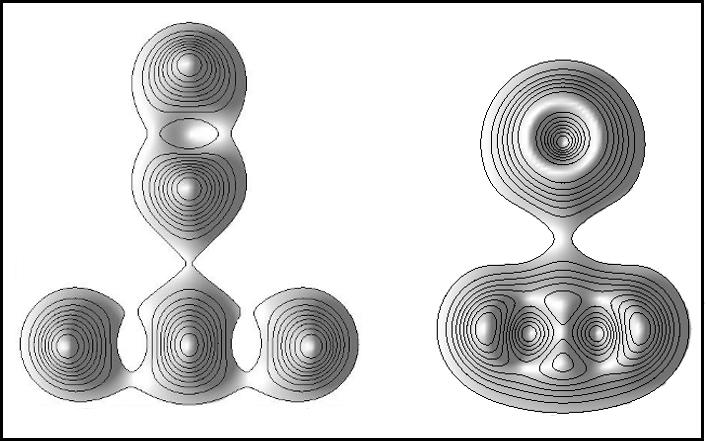}
\caption{
Two-dimensional plots of topological charge density, on the left five $B=1$ Baby Skyrmions, used as initial conditions to create the $B=3+2$ solution on the right.
}
\label{Q5-sym}

\end{figure}

A configuration of five randomly orientated $B=1$ Baby Skyrmions, equally spaced on a circle is used as the initial condition. This initial configuration gradient flows to a chain configuration, of energy $E_5 = 92.02$, differing from the $B=3+2$ solution proposed in \cite{BW1}. By setting up a system of Baby Skyrmions as in figure~\ref{Q5-sym}, which possesses an underlying symmetry about the $y$-axis, we are able to minimise to a $B=3+2$ solution, which has an energy of $E = 92.41$.

\begin{figure}[!ht]
 \centering
 \includegraphics[width = 11cm]{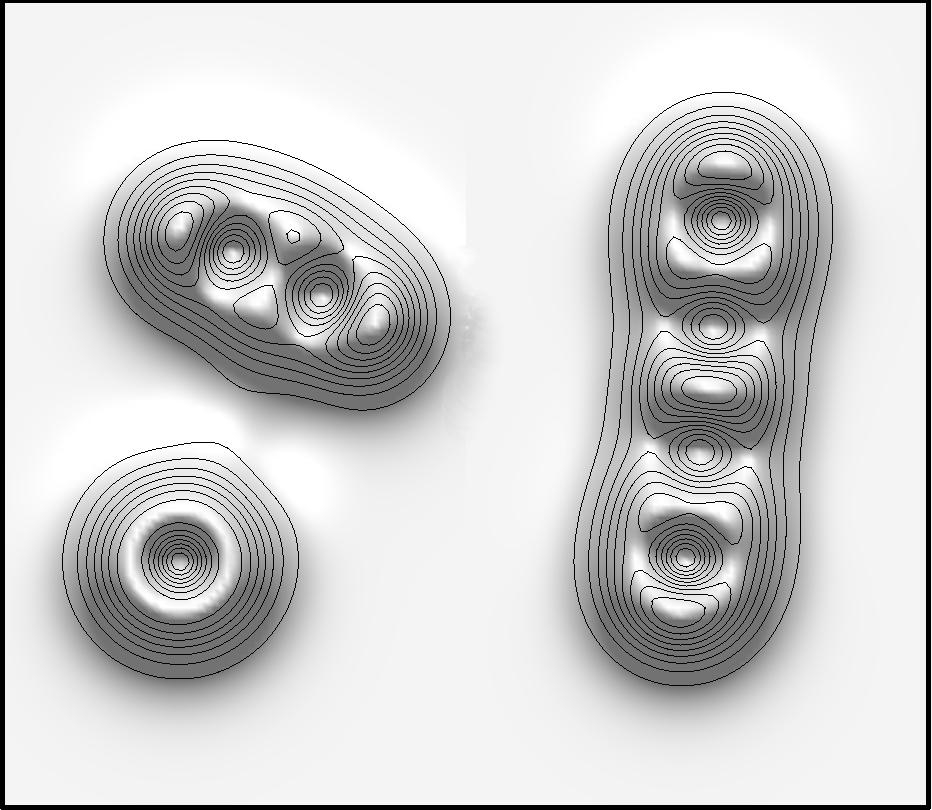}
\caption{
Two-dimensional plots of topological charge density of $B=5$ Baby Skyrmions,
with the previously accepted configuration of $B=3+2$ on the left which then flows to the new $B=5$ chain on the right.
}
\label{q5-both}

\end{figure}

Therefore we believe our chain solution is of a lower energy. A third justification of our results is shown in figure~\ref{q5-both}. Here it is clearly shown that the initial condition flows through a slightly perturbed $3+2$ solution, and minimises to the chain configuration. This shows that the chain solution must be of a lower energy than the accepted $3+2$ modular solution, which is a very long lived saddle point in the energetic landscape. Also, \cite{hen} has a $B=5$ static chain configuration for a potential term of the form $m^2(1-\phi_3)^{0.9}$, but there seems to be no comment on this new configuration.

\newpage

\subsection{$B=6$}

\begin{figure}[!ht]
 \centering
 \includegraphics[width = 8cm]{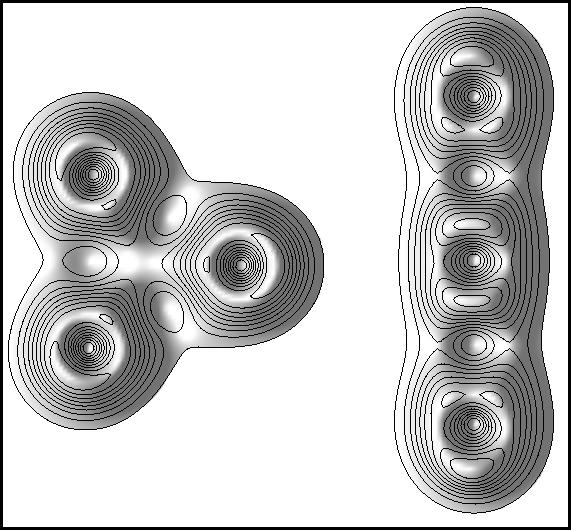}
\caption{
Two-dimensional plots of the topological charge densities of the previously accepted $B=2+2+2$ Baby Skyrmion triangular configuration on the left, and the new $B=6$ chain solution on the right.
}
\label{Q6-tri}

\end{figure}

The next configuration which differs from the common results of the literature is the $B=6$ case. Using a circle of six $B=1$ Baby Skyrmions as the initial configuration, led to a minimum energy configuration similar to the $B=6$ configuration presented in \cite{BW1}, a configuration of three $B=2$ Baby Skyrmions located at the vertices of an equilateral triangle as in figure \ref{Q6-tri}. We found the minimum energy of this triangular solution to be $E^{\Delta}_6 = 110.30$. This configuration does not match with the incremental structure common to all the previous lower charge configurations. This deviation from the chain form led us to minimise the energy of a row of six out of phase Baby Skyrmions, this produced a $B=6$ chain of static energy $E_6 = 110.22$. Due to the difference in energy being less than $0.1\%$, we believe the triangular structure to be a local minimum energy configuration.

\subsection{$B>6$}
For charges $6<B \leq 10$ we systematically found the minimum energy solution to be a Baby Skyrmion chain. We also energetically minimised some larger charges, namely $B=20,21$, and again found chain solutions.

\begin{figure}[!ht]
 \centering
 \includegraphics[width = 11cm]{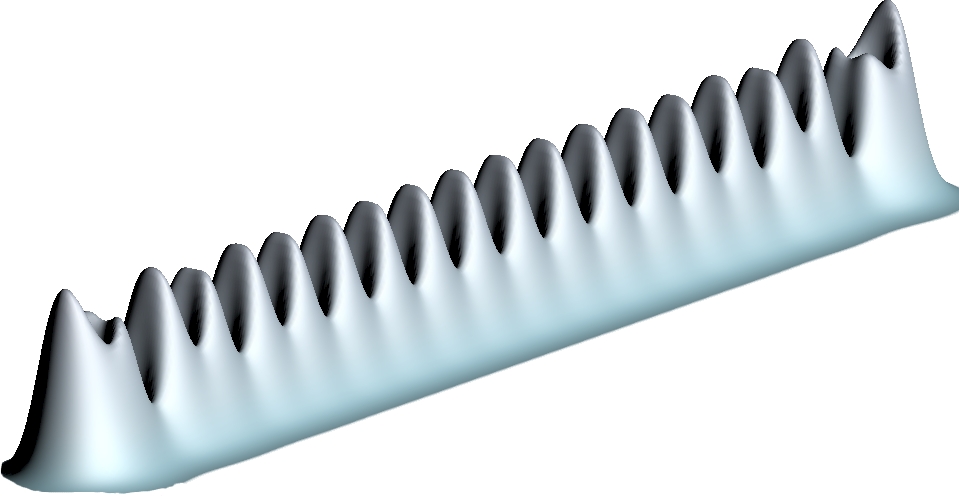}
\caption{
Two-dimensional plot of the topological charge density of the $B=20$ Baby Skyrmion chain.
}
\label{Q20-chain}

\end{figure}

\begin{figure}[!ht]
 \centering
 \includegraphics[width = 11cm]{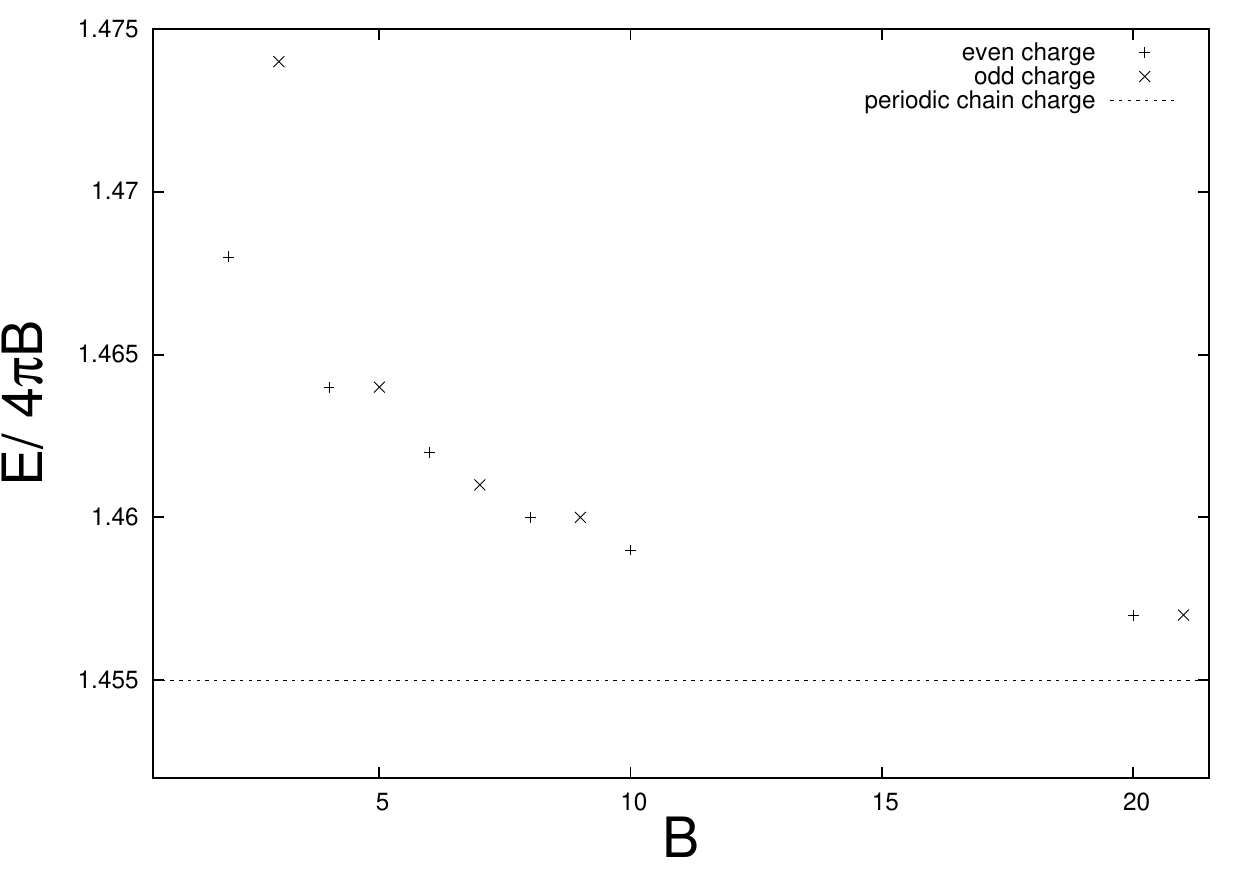}
\caption{
Plot of energy per-unit charge in units of $4\pi$, against the charge of the Baby Skyrmion chain.
}
\label{E-B}

\end{figure}

\begin{table}[!ht]
\centering
%\vskip -2cm
\begin{tabular}{|c|c|c|}
\hline
$B$ & $E$ & $E/4 \pi B$ \\
\hline
1 & 19.65 & 1.564 \\
2 & 36.90 & 1.468 \\
3 & 55.58 & 1.474 \\
4 & 73.61 & 1.464 \\
5 & 92.02 & 1.464 \\
6 & 110.22 & 1.462 \\
7 & 128.55 & 1.461 \\
8 & 146.81 & 1.460 \\
9 & 165.11 & 1.460 \\
10 & 183.38 & 1.459\\
20 & 366.20 & 1.457 \\
21 & 384.48 & 1.457 \\
\hline
\end{tabular}
\caption{Minimum energies of charged Baby Skyrmions
}
 \label{tab-energy}
\end{table}
 \newpage
Some features worth noting about the form of the chains, are that for large charge Baby Skyrmions the main central body of the chains appears to be made of uniform overlapped charge one Baby Skyrmions. Also the charge density tends to be less uniform nearer the ends of the chain, where upon they appear similar to highly perturbed charge two Baby Skyrmions. These $B=2$ Skyrmions seem to `cap' the ends of the chain, as in figure~\ref{Q20-chain} for the $B=20$. For small charges, the energy per charge does not decrease smoothly, as shown in figure ~\ref{E-B}.

Instead of a smooth decrease in energy for small $B$, the energy per-unit charge shows two trends, one for odd charge and one for even; these two trends then converge for increased $B$. This asymptotic energy per-unit charge and the periodic nature of the charge density of the central region of a long chain, as in figure ~\ref{Q20-chain}, motivated the calculation of the energy per-unit charge of an infinitely charged Baby Skyrmion chain, as described in the next section.

\section{Baby Skyrmions on a cylinder.} 
\begin{figure}[!ht]
 \centering
 \includegraphics[width = 11cm]{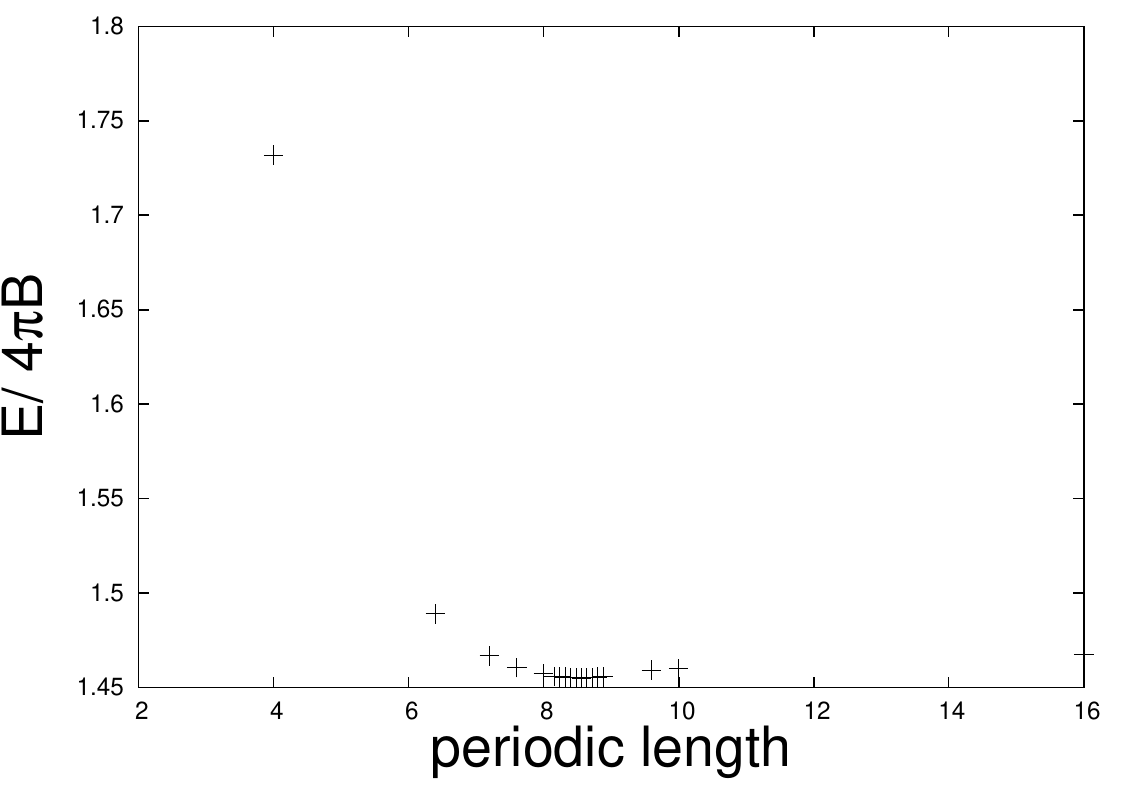}
\caption{
Plot of Baby Skyrmion chain energy per unit charge, in units of $4\pi$, for a varying periodic cell length.
}
\label{E-cell-cy}

\end{figure}

This section is similar to the work in \cite{DW}, with the aim to calculate the minimum Baby Skyrmion energy per charge of an infinitely charged chain. As presented in \cite{BW1}, it can be shown that two well separated Baby Skyrmions, with a relative phase $\Delta \chi$, have an interaction energy which can be calculated by a dipole approximation to be
\begin{equation}
 E_{int} = \frac{p^2 m^2}{\pi} K_0(m R)\cos (\Delta \chi), \label{int-eng}
\end{equation}
where $K_0$ is the order zero modified Bessel function, $p$ is a numerically found asymptotic decay constant and $R$ is the large separation distance. \eqref{int-eng} shows that the relative phase, $\Delta \chi$, of two Baby Skyrmions dictates the type of interaction. For $\Delta \chi =0$, the Baby Skyrmions repel, and for $\Delta \chi = \pi$ they attract.
Hence to create a stable chain of interacting Baby Skyrmions, we have to start with an anti-periodic chain of $B=1$ Baby Skyrmions. 
To proceed, the same numerical procedure as outlined above is implemented, but instead of imposing $\bphi = \mathbf{e}_3$ at the boundaries, we imposed periodic boundary conditions in the $y$-direction. We then vary the length of the periodic $y$-direction to minimise the energy as shown in figure \ref{E-cell-cy}. Due to the smaller lattice size than the infinite plane case, $\Delta x =0.08$ is used for a greater energy resolution.

This shows a minimum energy for a cell of periodic length $L=8.56$, for $\kappa =1 , m= \sqrt{0.1}$. The minimum energy for a Baby Skyrmion on a cylinder is found to be $E_{min} = 1.4549$, this energy is the dashed line plotted in \Fig ~\ref{E-B} which matches to the asymptotic of the energy-charge plot.

\section{Baby Skyrmion on a tours}
\begin{figure}[!ht]
 \centering
 \includegraphics[width = 11cm]{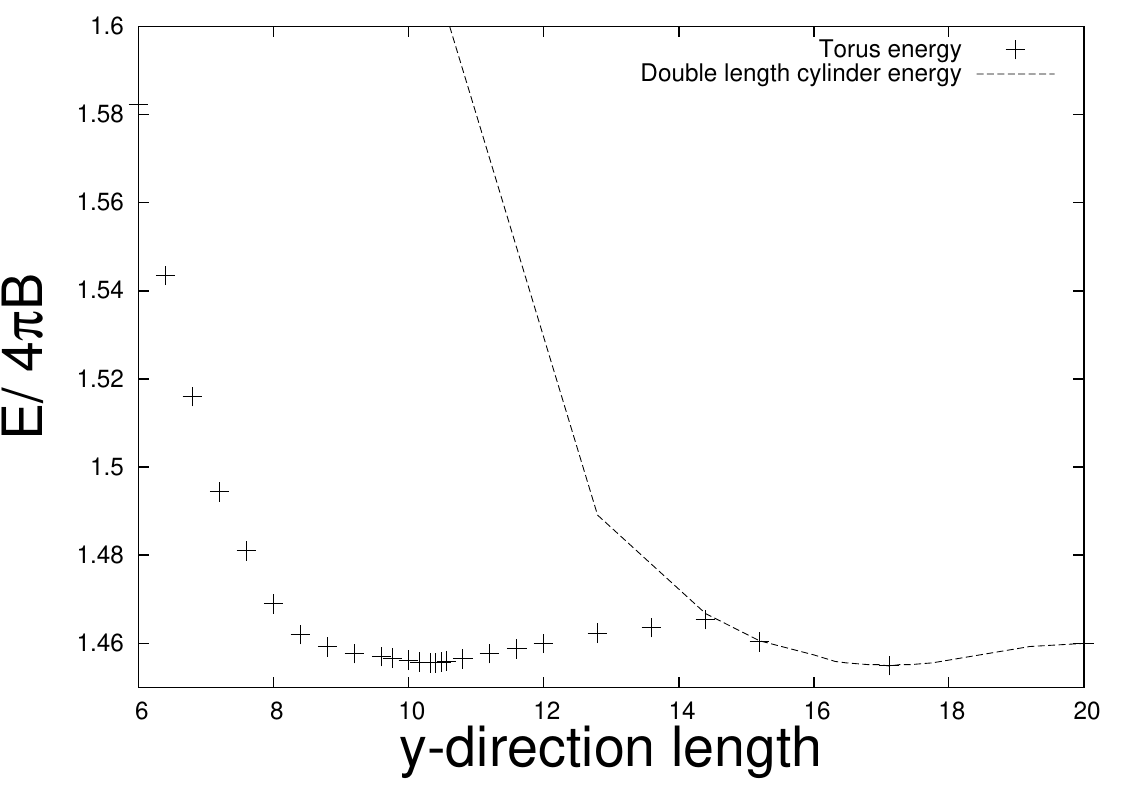}
\caption{
Plot of Baby Skyrmion chain energy per-unit charge, in units of $4 \pi$, for varying periodic cell length.
}
\label{E-cell-torus}

\end{figure}

With the aim to discover the minimum energy per charge of a Baby Skyrmion crystal, periodic boundary conditions need to be imposed. Hence the physical space is now a $2$-torus and so, $\bphi:\mathbb{T}^2 \to S^2$. \\
It is already known that the minimum energy configuration is that of a hexagonal lattice \cite{hen2}, where each hexagon hole effectively represents a charge of $B=1/2$. For numerical simplicity it is easier to work on a fundamental rectangular torus of size $ L \times \sqrt{3}L$, as in \cite{PP}. This lattice has the geometry to contain eight hexagons, hence a $B=4$ initial configuration is needed. To proceed we used the above algorithm to energetically minimise two $B=2$ Baby Skyrmions, on a $L \times \sqrt{3} L$ bi-periodic lattice. The length $L$ is then varied to find how the minimum energy varies with cell size, giving figure ~\ref{E-cell-torus}.

The result is that the natural unit cell size for a Baby Skyrmion lattice is a cell of size $L=10.4$, for $\kappa =1 , m=\sqrt{0.1}$. Figure~\ref{E-cell-torus} also has the energy per unit charge data for the cylinder case where the periodic length is doubled to allow for a $B=4$ unit cell. This shows that as the fundamental rectangle grows in size, locally the Baby Skyrmion behave as if it were on a cylinder. So the Skymion configuration for increasing fundamental rectangle, first starts off as a compressed crystal, then becomes an hexagonal crystal, then a deformed chain, then a energetically minimum chain and finally a single $B=4$ Baby Skymion on an infinite plane. One of the main points of interest of the torus case, is that the minimum energy of a lattice $E_{latt} = 1.4555$ and the minimum $B=4$ chain $E_{chain}=1.4549$ are well within $ 0.05 \%$. This difference in energy is smaller than the numerical accuracy hence it cannot be concluded which, if either, is the minimum energy solution, only that the two configurations have very comparable energies.

\section{Conclusion}
We have presented numerical solutions for multi-charge Baby Skyrmions which differ to the configurations in the literature. This discrepancy is believed to be due to the extra computing power now available, which allowed us to find lower energy solutions for $B \geq 5$. This has led to the proposal that Baby Skymion minimum energy configurations are chains, and we are confident in this result because of the smooth lowering of energy per charge for increasing charge configurations. Also, the energy per-unit charge asymptotes to the periodic chain energy per-unit charge case, showing the increasing stability as the charge of a chain grows. \\
The planar hexagonal lattice and the periodic chains minimum energies are too similar to be able to identify the preferred configuration. This energetic similarity between the hexagonal lattice and the chain energies is similar to the negligible difference between the commonly accepted $B=2+2+2$ solution and the $B=6$ chain solution. The energetic similarity between the $B=2+2+2$ and the $B=6$ chain is an exact analogue to the full Skyrme model, where it is found that the $B=12$ solution can be either a linear chain of three $B=4$ Skyrmions, or a configuration of three $B=4$ Skyrmions, each placed at a vertex of an equilateral triangle \cite{Sk-alp}. These two configurations, again have similar energies.

An interesting extension to this work would be to extend the numerical algorithms to higher orders, with the aim to be able to confidently identify the minimum energy periodic charge configuration. Also it would be interesting to see if similar chain solutions are also minimum energy solutions to the full Skyrme model.

\section{Acknowledgements}
I thank STFC for a research studentship and B.M.A.G. ~ Piette, W.J ~Zakrzewski and P.~Sutcliffe for their extremely constructive advice.

\end{document}